\documentclass[intlimits,twoside,a4paper]{article}

\usepackage{amsmath,amssymb}
\usepackage{graphicx}
\usepackage[T2A]{fontenc}
\usepackage[cp1251]{inputenc}
\usepackage{graphics}  

\newcommand{\bea}{\begin{eqnarray}}
\newcommand{\eea}{\end{eqnarray}}
\newcommand{\ba}{\begin{array}}
\newcommand{\ea}{\end{array}}
\newcommand{\edc}{\end{document}}
\newcommand{\bc}{\begin{center}}
\newcommand{\ec}{\end{center}}
\newcommand{\be}{\begin{equation}}
\newcommand{\ee}{\end{equation}}

\def\bc{{\mathbb C}}

\def\g{\gamma}

\newtheorem{thm}{Theorem}[section]

\newtheorem{prop}[thm]{Proposition}
\newtheorem{defin}[thm]{Definition}
\newtheorem{rem}{Remark}[section]
\newtheorem{ex}{Example}[section]

\DeclareMathOperator{\Fix}{Fix}

\usepackage[eqsecnum]{cmpj3}

\issue{2019}{22}{2}{23002}
\doinumber{10.5488/CMP.22.23002}

\title[Gibbs measures on triangle chandelier-lattices]
{Gibbs measures of an Ising model with competing interactions on the triangular chandelier-lattice}
\author[H. Ak\i n]{H. Ak\i n\thanks{akinhasan25@gmail.com}}
\address{Ceyhun Atuf Kansu Caddesi 1164. Sokak, 9/4, TR06105, \c{C}ankaya, Ankara, Turkey}

\date{Received May 3, 2019, in final form June 19, 2019}

\begin{document}

\maketitle

\begin{abstract}
In this paper, we consider an Ising model with three competing
interactions on a triangular chandelier-lattice (TCL). We describe
the existence, uniqueness, and non-uniqueness of
translation-invariant Gibbs measures associated with the Ising
model. We obtain an explicit formula for Gibbs measures with a
memory of length 2 satisfying consistency conditions. It is rigorously proved
 that the model exhibits phase transitions only for
given values of the coupling constants. As a consequence of our
approach, the dichotomy between alternative solutions of
Hamiltonian models on TCLs is solved. Finally, two numerical
examples are given to illustrate the usefulness and effectiveness
of the proposed theoretical results.

\pacs 05.70.Fh,  05.70.Ce, 75.10.Hk
\end{abstract}

\section{Introduction}

As is known, Cayley tree (or Bethe lattice), introduced by Hans
Bethe in 1935, is a non-realistic lattice. Since other operations
and the calculations on this lattice are easier to understand than
the $d$-dimensional $\mathbf{Z}^{d}$ lattice, many of the topics
in statistical physics have recently been taken into account on
the Cayley tree \cite{Akin2016}. Thus, the results obtained on the
Cayley tree became a source of inspiration for the $d$-dimensional
$\mathbf{Z}^{d}$ lattice. As a result, many researchers have
employed the Ising and Potts models \cite{GTA,AkinT2011CMP} in
conjunction with the Cayley tree \cite{BRZ,GTA,GATTMJ,UGAT}. The
Ising model has relevance to physical, chemical, and biological
systems \cite{Lebowitz1,G,Iof}.

We were then able to identify a similar lattice, which we
identified as triple, quadruple, quintuple, and so on. So we
examined the dynamic behaviours of Ising models on these
Cayley-like lattices. Up till now, some studies have been done
\cite{AUT2010AIP,UA2011CJP,UA2010PhysicaA,UGAT2012ACTA}. Although
the results are similar to the results for the models on the
Cayley tree, we think that we would get many different results in
the future. We called this model a triangular, rectangular,
pentagonal and similar ``chandelier'' model. Compared with
$\mathbf{Z}^{d}$ lattice, we think that the chandelier lattice is
more realistic than the Cayley tree
\cite{AUT2010AIP,UA2011CJP,UA2010PhysicaA,Moraal,UGAT2012ACTA}.
In this paper, we deal with a Cayley tree-like lattice
\cite{UA2010PhysicaA} which we called a \textbf{triangular
chandelier lattice} (shortly, TCL) from the configuration model.

The theory of probability is one of the basic branches of
mathematics lying at the base of the theory of statistical
mechanics \cite{G,ART,Bax,D1,Preston,Zachary,Sinai}. As is known,
one of the fundamental problems of statistical mechanics is to
specify the set of all Gibbs measures associated to the given
Hamiltonian \cite{NHSS,NHSS1,RAU,GHRR,AGTU2013ACTA}. A Gibbs
measure is a probability measure frequently used in many problems
of probability theory and statistical mechanics. It is also known
that such measures form a convex compact subset that is different
from the void in the set of all probability measures. The number
of translation-invariant splitting Gibbs measures associated with
the Ising model on a Cayley tree can only be one or more than one,
depending on temperature \cite{Utkir-GM2013}. The $p$-adic
counterpart of the Ising-Vanniminus model on the Cayley tree of
order two was first studied in \cite{MDA}. There was proposed
a measure-theoretical approach to investigate the model in the
$p$-adic setting.

In this present paper, we want to investigate
translation-invariant Gibbs measures (TIGMs) corresponding to an
Ising model on the TCL. It is well known that the comprehension of
phase transitions is one of the most interesting, perhaps the
central, problems of equilibrium statistical mechanics
\cite{Lebowitz1}. By the phase transition we mean the existence of
at least two distinct Gibbs measures associated with the given
model \cite{Akin2016,Akin2017,G,Preston}. We will investigate the
existence of translation invariant Gibbs measures on a wide class
of the TCL, restricted only to the memory of length 2. We derive
specific realizations for the ANNNI model on these structures. We
derive the results within the Markov random field framework,
making use of the Kolmogorov consistency conditions. We express
the solutions of recurrence relations warranting consistency in
terms of the fixed points of a function $f(x)$. We provide some
diagrams of the behaviour of the function $f(x)$ for different
values of the model parameters. We present analytical developments
allowing for the identification of Gibbs measures along usual
procedures.

The structure of the present article is as follows: in section
\ref{Preliminary}, we give the necessary definition and
preliminaries about Ising model with three competing interactions
on a TCL. In section \ref{Gibbs Measures}, we establish the Gibbs
measure associated with the model. In section \ref{Section-TIGMs},
we describe the existence, uniqueness and non-uniqueness of
translation-invariant Gibbs measures associated with the Ising
model on a TCL. In section \ref{Phase translations}, it is rigorously proved
 that the model exhibits phase transitions only for
given values of the coupling constants. As a consequence of our
approach, the dichotomy between alternative solutions of
Hamiltonian models on TCLs is solved. Finally, in section
\ref{Conclusions}, the relevance of the results obtained for
systems on the TCL is discussed and the results are compared to
ones on the Cayley tree.

\section{Preliminary}\label{Preliminary}
\subsection{Triangular chandelier lattice}
\begin{figure}[!b]
\centering
\includegraphics[width=65mm]{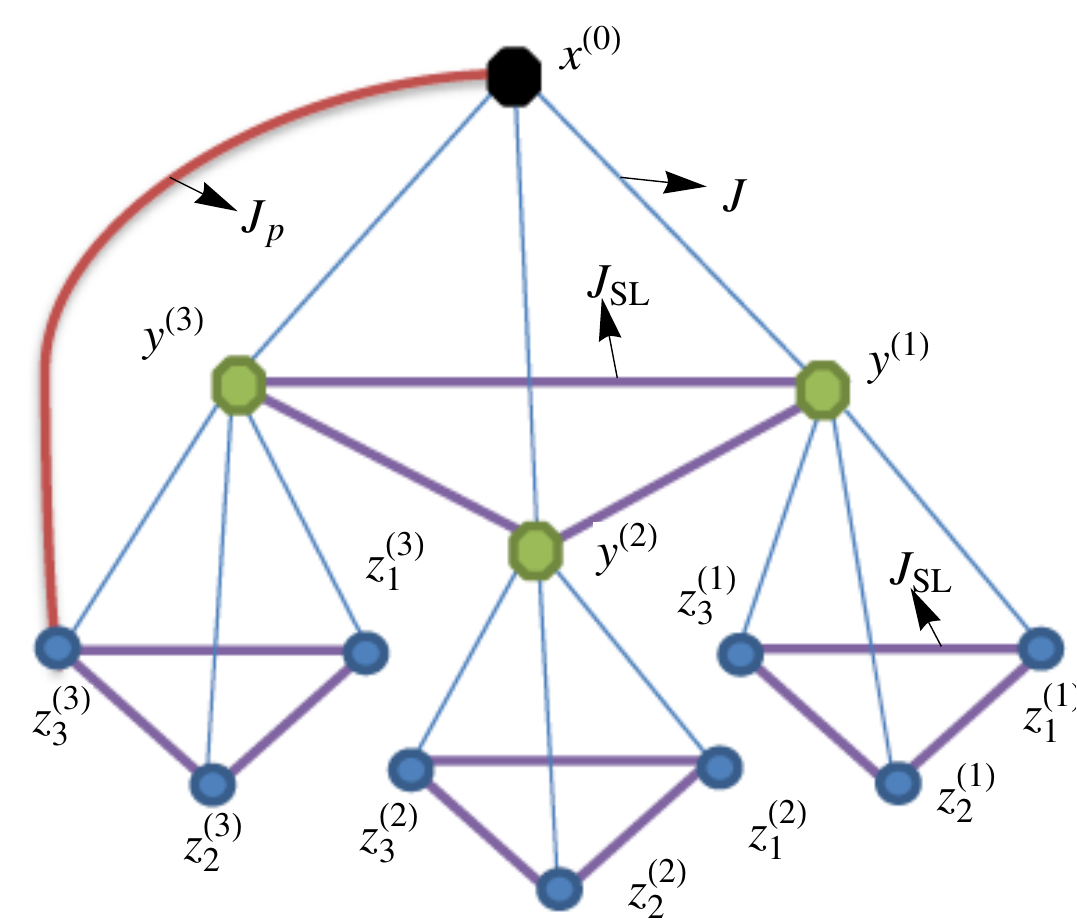}
\caption{(Colour online) Cayley tree-like lattice: triangular
chandelier with 2 level. Three successive generations of TCL ($J$
represents nearest-neighbour interactions; $J_\text{p}$ represents
prolonged next nearest-neighbour interactions and $J_\text{SL}$
represents same-level nearest-neighbours interactions.
}\label{ternary-chandiler}
\end{figure}

Chandelier lattices are simple connected undirected graphs $G =
(V, E)$ ($V$ set of vertices, $E$ set of edges).
Let $C^k=(V, E, i)$ be order $k$ chandelier lattice  with a root
vertex $x^{(0)}\in V$, where each vertex has $(k + 3)$ nearest
neighbours with $V$ as the set of vertices and the set of edges. It
is clear that the root vertex $x^{(0)}$ has $k$ nearest neighbours.
The notation $i$ represents the incidence function corresponding
to each edge $e\in E$, with end points $x_1,x_2\in V$. There is a
distance $d(x, y)$ on $V$, the length of the minimal point from $x$
to $y$, with the assumed length of 1 for any edge (see figure~\ref{ternary-chandiler}).

Let us consider a chandelier with 3 lamps hanging on the
ceiling. Suppose that the same three quadrants hanging on each lamp
of the first chandelier were added. In this case, we get a weave
that resembles a semi-infinite Cayley tree. We assume here that
each lamp is connected to the lamps in the nearest neighbours.
Thus, we can have the possibility to investigate the titles
examined in statistical physics by calculating the internal,
external and full energies corresponding to a Hamiltonian on the
chandelier lattice that we have defined.

The distance $d(x,y)$, $x,y\in V$, on the chandelier lattice $C^k$
($k>2$), is the number of edges in the shortest path from $x$ to
$y$. The fixed vertex $x^{(0)}$ is called the $0$-th level and the
vertices in $W_n$ are called the $n$-th level. For the sake of
simplicity we put $|x|=d(x,x^{(0)})$, $x\in V$. We denote the
sphere of radius $n$ on $V$ by
\[
W_n^{(P)}=\{x\in V: d(x,x^{(0)})=n \}
\]
and the ball of radius $n$ by
\[
V_n^{(P)}=\{x\in V: d(x,x^{(0)})\leqslant n \},
\]
where vertex $x$ is prolonged downwards relative to $x^{(0)}$.
\[
L_n=\{l=\langle x, y\rangle\in L |x, y\in V_n\}.
\]
For example, $W_2^{(P)}=\{z^{(u)}_v: u,v=1,2,3 \}$ (see figure~\ref{ternary-chandiler}).

The set of direct prolonged successors of any vertex $x\in W_n$ is
denoted by
\[
S_k^{(P)}(x)=\{y\in W_{n+1}:d(x,y)=1 \}.
\]
The set of same-level neighbourhoods of any vertex $x\in W_n$ will
be denoted by
\[
SL_k(x)=\{y\in W_{n}:d(x,y)=1 \}.
\]
It is clear that $|SL_k(x)|=2$, for all vertexes $x\in W_n$.
\begin{defin}\label{neighbourhoods}
Hereafter, we will use the following definitions for
neighbourhoods.
\begin{enumerate}
    \item Two vertices $x$ and $y$, $x,y \in V$ are called {\it
\textbf{nearest-neighbours (NN)}} if there exists an edge $e\in E$
connecting them, which is denoted by $e=\langle x,y\rangle$.
    \item The nearest-neighbour vertices $x,y\in V$ that are not
prolonged are called {\it \textbf{same-level nearest-neighbours
(SLNN)}} if $|x|=|y|$ and are denoted by $\widetilde{\rangle x,y\langle}$.
    \item Two vertices $x,y\in V$ are called {\it \textbf{the
next-nearest-neighbours (NNN)}} if there exists a vertex $z \in V$
such that $x, z$ and $y, z$ are NN, that is if $d(x,y)=2$.
    \item The next-nearest-neighbour
vertices $x\in W_n$ and $y\in W_{n+2}$ are called {\it
\textbf{prolonged next-nearest-neighbours (PNNN)}} if $|x|\neq |y|$
and is denoted by $\rangle x,y\langle$ (see figure~\ref{ternary-chandiler}).
\end{enumerate}
\end{defin}

\subsection{Kolmogorov consistency condition}
Kolmogorov's extension theorem allows us to construct a
variety of measures on infinite-dimensional spaces (see
\cite{Cinlar} for details).

Now, let us explain this theorem for a one-dimensional situation.
Let $S=\{0,2,\ldots,k-1\}$ be a finite state space. On the infinite
product space $\Omega={{S}^{\mathbf{Z}}}$, one can define the
product $\sigma$-algebra, which is generated by cylinder sets
$_{m}[i_1,\ldots,i_N]=\{x\in S^{\mathbf{Z}}:x_{m}=i_0,\ldots ,
x_{m+N-1}=i_N\}$  of length $N$ based on the block $(i_1,\ldots,i_N)$
at the place $m$. Note that a cylinder set is a set of sequences
where we fix which symbol can occur in a finite number of places.
We denote by ${{\mathfrak{M}}}({{S}^{\mathbf{Z}}})$ the set of all
measures on ${{S}^{\mathbf{Z}}}$. The set of all
$\sigma$-invariant measures in ${{S}^{\mathbf{Z}}}$
is denoted by ${{\mathfrak{M}}_{\sigma }}({{S}^{\mathbf{Z}}})$,
where $\sigma$ is the shift transformation.
\begin{prop}\cite[(8.1) Proposition]{Denker}\label{Kolmo-Cons1}
For $\mu \in {{\mathfrak{M}}_{\sigma }}({{S}^{\mathbf{Z}}})$, the
following properties are valid:
\begin{enumerate}
    \item $\sum\limits_{i\in S}{\mu {{(}_{0}}[i])=1}$;
    \item $\mu (_{n}[i_{0},\ldots,i_{k}])\geqslant 0$ for any block
$({{i}_{0}},{{i}_{1}},\ldots,{{i}_{k}})\in S^{k+1}$ and any $n\in
\mathbf{Z}$;
    \item $\mu (_{n}[{{i}_{0}},\ldots,{{i}_{k}}])=\sum\limits_{{{i}_{k+1}}\in S}{\mu
    {{(}_{n}}[{{i}_{0}},\ldots,{{i}_{k}},{{i}_{k+1}}])}$;
    \item  $\mu (_{n}[{{i}_{0}},\ldots,{{i}_{k}}])=\sum\limits_{{{i}_{-1}}\in S}{\mu{{(}_{n}}[{{i}_{-1}},{{i}_{0}},\ldots,{{i}_{k}}])}$.
\end{enumerate}
\end{prop}
By a special case of Kolmogorov's consistency theorem (see
\cite{Denker}), these properties are sufficient to define a
measure. It is well known that a Gibbs measure is a generalization
of a Markov measure to any graph. Therefore, any Gibbs measure
should satisfy the conditions in the proposition
\ref{Kolmo-Cons1}.

We shall give two examples satisfying the conditions in proposition
\ref{Kolmo-Cons1} and to illustrate the consistency conditions.
\begin{ex}\label{Bernoulli}
Let $\pi=(p_i)_{i\in S}$ be any probability
vector on the state set $S$. For each $n\geqslant 0$, define
\begin{equation}\label{Ber-Measure}
\mu_{\pi}(_{m}[{{i}_{0}},\ldots,{{i}_{n}}])=p_n(i_0,i_1,\ldots,i_n)=p_{i_0}p_{i_1}\ldots
p_{i_n},
\end{equation}
where $i_0,i_1,\ldots,i_n \in S$. It is clear that $\{p_n\}_{n\geqslant
0}$ satisfies the consistency conditions (1)--(4) in proposition~\ref{Kolmo-Cons1} (see \cite{Denker}). Such a measure $\mu_{\pi}$
is called a \textbf{Bernoulli measure}. One of motivation examples
is the Bernoulli measure, which also satisfies the compatible
property.
\end{ex}
\begin{ex}\label{Markov}
Let $\pi=(p_i)_{i\in S}$ be any probability vector on the state
set $S$ and let $P=(p_{ij})_{i,j\in S}$ be any stochastic matrix,
i.e., $ 0\leqslant p_{ij}\leqslant 1$ and $\sum_{k \in S}p_{ik}=1$ for each
$i,j\in S.$ Thus, $\pi$ is defined as a probability vector such
that $\pi P =\pi$. If $P$ is irreducible, $\pi$ is uniquely
defined.

For each $n\geqslant 0$, the function defined by
\[
\mu_{\pi P}
(_{m}[i_{0},\ldots,i_{n}])=p_n(i_0,i_1,\ldots,i_n)=p_{i_0}p_{i_0i_1}\ldots
p_{i_{n-1}i_n},\ \ \ \ \text{where} \ \ i_0,i_1,\ldots,i_n \in S
\]
satisfies the consistency conditions (1)--(4) in proposition
\ref{Kolmo-Cons1} (see \cite{Denker}). Such a measure $\mu_{\pi
P}$ is called a \textbf{Markov measure}.
\end{ex}

The proof of the proposition \ref{Kolmo-Cons1} can clearly be
checked for both the Bernoulli and the Markov measures on
$\sigma$-algebra \cite{Denker}. For any cylinder set
$_{m}[i_0,\ldots,i_n]=\{x\in S^{\mathbf{Z}}:x_{m}=i_0,\ldots ,
x_{m+n-1}=i_n\}$ and any $k\geqslant 1,$ we have
\[
_{m}[i_0,\ldots,i_n]=\bigcup_{i_{n+1}\in S}\ldots\bigcup _{i_{n+k}\in
S}(_m[i_0,\ldots,i_n,i_{n+1},\ldots,i_{n+k}])
\]
and
\[
p_n(_{m}[i_0,\ldots,i_n])=\sum _{i_{n+1}\in S} \text{\ldots}\sum
_{i_{n+k}\in S}
p_{n+k}(_m[i_0,\text{\ldots},i_n,i_{n+1},\text{\ldots},i_{n+k}]).
\]
The following is the Kolmogorov extension theorem.
\begin{thm}\cite[4.18 Theorem]{Cinlar}\label{KolET1} Let
$S=\{0,1,\ldots,r-1\}$, for some $r\geqslant 2.$ Let $\{p_n\}_{n\geqslant 0}$
be a sequence of functions satisfying the consistency conditions,
where $p_n$ has domain $S^{n+1}$. Then, there exists a unique
probability measure $\mu$ on the measurable space $(\Omega,
B(\Omega))$ such that
\[
\mu (_{m}[i_{0},\ldots,i_{n}])=p_n(i_0,i_1,\ldots,i_n)
\]
for all $i_0,i_1,\ldots,i_n \in S$ and all $n\geqslant 0$.
\end{thm}
\section{Gibbs measures}\label{Gibbs Measures}
Let us consider the Ising  model with competing nearest-neighbour
interactions defined by the Hamiltonian
\begin{equation}\label{hm}
H(\sigma)=-J\sum_{\langle x,y\rangle\in L_n}\sigma(x)\sigma(y),
\end{equation}
where the sum runs over nearest-neighbour vertices $\langle x,y\rangle$ and the
spins $\sigma(x)$ and $\sigma(y)$ take values in the set
$\Phi=\{-1,+1\}$.

Let $h_x$ be a real-valued function of $x\in V$. A
finite-dimensional Gibbs distributions on $\Phi^{V_n}$ are defined
by formula
\begin{equation}\label{mu1}
\mu_n(\sigma_n)=\frac{1}{Z_{n}}\exp\left[-\frac{1}{T}H_n(\sigma_n)+\sum_{x\in
W_{n}}\sigma(x)h_{x}\right]
\end{equation}
with the associated partition function defined as
\begin{equation*}\label{partition1}
Z_n=\sum_{\sigma_n \in
\Phi^{V_n}}\exp\left[-\frac{1}{T}H_n(\sigma)+\sum_{x\in
W_{n}}\sigma(x)h_{x}\right],
\end{equation*}
where the spin configurations $\sigma_n$ belong to $\Phi^{V_n}$
and $h=\{h_x\in \mathbf{R},x\in V\}$ is a collection of real
numbers that define boundary condition (see \cite{BleherG}). This
distribution is a measure \cite{Bleher-Zalys,Bleher1990a}.

Bleher and Zalys \cite{Bleher-Zalys} studied the existence of
limit distributions for the ferromagnetic Ising model on infinite
diamond-shaped hierarchical lattice (DHL). They have proved that
for low temperatures and zero external field, there exist exactly
two extreme Gibbs limit distributions, and in other cases the
Gibbs distribution is unique.

We say that the probability distributions \eqref{mu1} are
compatible if for all $n\geqslant 1$ and $\sigma_{n-1}\in
\Phi^{V_{n-1}}$:
\begin{equation}\label{Gibbs-KCC}
\sum _{\omega _n\in \Phi ^{W_n}} \mu _n(\sigma_{n-1}\lor \omega
_n)=\mu_{n-1}(\sigma_{n-1}).
\end{equation}
Here, $\sigma _{n-1}\lor \omega_n$ is the concatenation of the
configurations. It is clear that the equation \eqref{Gibbs-KCC} is
the same as condition (4) in proposition \ref{Kolmo-Cons1}.

In this case, according to theorem \ref{KolET1} (the Kolmogorov
theorem), there exists a unique measure $\mu$ on $\Phi^V$ such
that, for all $n$ and $\sigma_n\in \Phi^{V_n}$
\[
\mu \left(\left\{\sigma |_{V_n}=\sigma _n\right\}\right)=\mu _n(\sigma _n).
\]
Such a measure is called a \textbf{splitting Gibbs measure} with
memory of length 1 corresponding to the Hamiltonian \eqref{hm} and
function $h_x$, $x\in V$ (see \cite{Utkir-GM2013} for details).

Previously, researchers frequently used memory of length 1 over a
Cayley tree to study Gibbs measures \cite{BleherG}. In
\cite{Akin2016,NHSS,NHSS1,MAKfree2017}, the authors have studied
Gibbs measures with a memory of length 2 for generalized ANNNI
models on a Cayley tree of order 2 by means of a vector valued
function (see \cite{FV} for details). In \cite{Akin2016,NHSS}, the
next generalizations are considered. These authors have defined
Gibbs measures or Gibbs states with a memory of length 2 (on
spin-configurations $\sigma$) for generalized ANNNI models on
Cayley trees of order 2. In \cite{MAKfree2017}, the authors have
obtained some rigorous results to propose a measure-theoretical
approach for the Ising-Vannimenus model.

\subsection{New Gibbs measures}
In this subsection we are going to construct new Gibbs measures
associated with the Ising model on the TCL. We consider the
following Hamiltonian
\begin{equation}\label{Hm-C3}
H(\sigma )=-J\sum\limits_{\langle x,y\rangle}{\sigma (x)\sigma
(y)}-{{J}_\text{p}}\sum\limits_{\rangle x,y\langle}{\sigma (x)\sigma
(y)}-{{J}_\text{SL}}\sum\limits_{\widetilde{\langle x,y\rangle}}{\sigma (x)\sigma
(y)},
\end{equation}
where the first sum ranges all nearest neighbours, the sum in the
second term ranges all prolonged next-nearest-neighbours, the third
sum ranges all same-level nearest-neighbours and $J,
J_\text{p},J_\text{SL}\in \mathbf{R}$ are coupling constants (see figure~\ref{ternary-chandiler}). Note that if $J_\text{SL}=0$, then the
Hamiltonian \eqref{Hm-C3} coincides with the Vannimenus's
Hamiltonian \cite{Vannimenus}.

Let us introduce a class of Markov chains on the TCL $C^3$. For
$\Lambda \subset V$ denote $\Phi^{\Lambda}=\{ - 1, +
1\}^{\Lambda}$, the configurational space of the set $\Lambda$.
For a finite subset $V_n$ of TCL, we define the finite-dimensional
Gibbs probability distributions on the configuration space
\[
\Omega^{V_n}=\{\sigma_n=\{\sigma(x)=\pm 1, x\in V_n \}\}
\]
at inverse temperature $\beta=\frac{1}{kT}$ by formula
\begin{equation}\label{mu}
\mu_{\textbf{h}}^{(n)}(\sigma_n) =\frac{1}{Z_{n}}\exp\bigg[-\beta
H_n(\sigma_n)+\sum_{
%\begin{array}{c}
 x\in W_{n-1}
%\end{array}
}\sigma(x)\sigma(y)\sigma(z)\sigma(w)h_{B_1(x),\sigma(x)\sigma(y)\sigma(z)\sigma(w)}\bigg]
\end{equation}
with the corresponding partition function defined by
\begin{equation*}
Z_n=\sum\limits_{\sigma_n\in \Omega^{V_n}}\exp\bigg[-\beta
H_n(\sigma_n)+\sum_{
%\begin{array}{c}
 x\in W_{n-1}
%\end{array}
}\sigma(x)\sigma(y)\sigma(z)\sigma(w)h_{B_1(x),\sigma(x)\sigma(y)\sigma(z)\sigma(w)}\bigg],
\end{equation*}
where $y,z,w\in S^{(P)}(x)$.

Let us give the construction of a special class of limiting Gibbs
measures associated with the Ising model corresponding to the
Hamiltonian \eqref{Hm-C3} on the TCL. Firstly, we show that the
Gibbs measures associated with the Ising model satisfy the
conditions in the proposition \ref{Kolmo-Cons1}.

The following statement describes conditions on
$h_{B_1(x),\sigma(x)\sigma(y)\sigma(z)\sigma(w)}$ ensuring
compatibility of $\mu_{\textbf{h}}^{(n)}$.

\begin{thm}\label{Compatible-thm1}
Probability distributions $\mu_{\textbf{\textup{h}}}^{(n)}$, $n = 1, 2,
\ldots \,$, in \eqref{mu} are compatible if for any $x\in V$ the
following equation holds:
\begin{align}\label{compatible2a}\nonumber
&\exp\bigg[{\sigma(x)\prod_{v=1}^3
\sigma(y^{(v)})h_{B_1(x),S_{3-i}^i(\sigma (x))}}\bigg]=L_2\sum _{
\substack{ y^{(v)}\in S(x)\\
z_u^{(v)}\in S(y^{(v)})}
}\Bigg\{\exp\bigg[{\beta J\sigma (y^{(v)})\sum _{u=1}^3 \eta
(z_u^{(v)})}\bigg]\\
&\times\exp\bigg[{\beta J_\text{\textup{p}}\sigma (x)\sum _{u,v=1}^3 \eta
(z_u^{(v)})}\bigg]
 \exp\bigg[\beta J_\text{\textup{SL}}\Big\{\sigma
(z_1^{(v)})\sigma(z_2^{(v)})+\sigma(z_3^{(v)})\big[\sigma
(z_1^{(v)})+\sigma(z_2^{(v)})\big]\Big\}\bigg]\nonumber\\  &\times \prod _{v=1}^3
\exp\bigg[{\sigma (y^{(v)})\prod _{u=1}^3 \eta
(z_u^{(v)})h_{B_1(y^{(v)}),S_{3-i}^i(\sigma (y^{(v)}))}}\bigg]\Bigg\}.
\end{align}
Here, $S(x)$ is the set of direct prolonged  successors of $x$ on
TCL, and $S(y^{(v)})$ is the set of direct prolonged successors of
$y^{(v)}$ on TCL and $L_2=\frac{Z_1}{Z_2}$.
\end{thm}
The proof can be done similarly to \cite{Akin2017}.

\subsection{Basic equations}

\begin{figure}[!t]
\centering
\includegraphics[width=35mm]{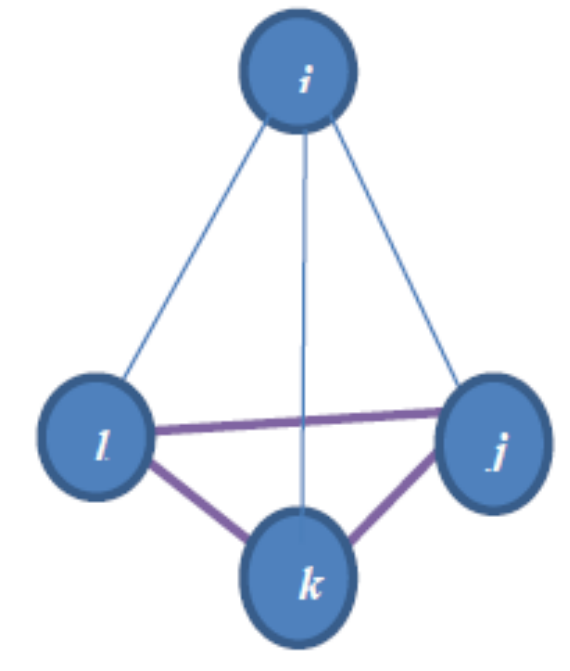}
\caption{(Colour online) Possible configurations on unit semi-ball
$B_1(x)$ on the TCL of order three ($i,j,k,l\in
\{-,+\})$.}\label{TC1l}
\end{figure}

Denote $B_1(x)=\{x,y,z,w\}$ a unit semi-ball with a center $x$,
where $S(x)=\{y,z,w\}$. For the sake of simplicity, from figure~\ref{ternary-chandiler}, we assume that
\[
B_1(x^{(0)})=\{x^{(0)},y^{(1)},y^{(2)},y^{(3)}\},\quad
B_1(y^{(i)})=\{y^{(i)},z^{(i)}_1,z^{(i)}_2,z^{(i)}_3\} \quad \text{for} \ \
i=1,2,3.
\]
Let $x\in W_{n}^{(P)}$ for some $n$ and $S(x)=\{y,z,w\}$, where
$y,z,w\in W_{n+1}^{(P)}$ are the direct successors of $x$.
We denote the set of all spin configurations on $V_n^{(P)}$ by
$\Phi^{V_n^{(P)}}$ and the set of all configurations on unit
semi-ball $B_1(x)$ by $\Phi^{B_1(x)}$ (see figure~\ref{TC1l}). One
can get that the set $\Phi^{B_1(x)}$ consists of  sixteen
configurations
\begin{equation}\label{config1}
\Phi^{B_1(x)}=\left\{\left(
  \begin{array}{ccc}
      & i & \\
      &  & \\
     l & & j\\
     & k &
  \end{array}
\right)=( ijkl): i,j,k,l\in \{-1,+1\} \right\}.
\end{equation}
For example, we
have \[S^{0}_3(+)=\sigma_1^{(1)}=\left(
  \begin{array}{ccc}
      & + & \\
      &  & \\
     + & & +\\
     & + &
  \end{array}
\right)=(+++\,+).\] For the sake of simplicity, let us consider the
following abbreviations:
\begin{eqnarray*}\label{h-funct}
&&h_1=h_{B_1(x),\sigma_1^{(1)}}=h_{B_1(x),S^{0}_3(+)}=h_{B_1(x),++++},\\
&&h_2=h_{B_1(x),\sigma_2^{(1)}}=h_{B_1(x),\sigma_3^{(1)}}=h_{B_1(x),\sigma_4^{(1)}}=h_{B_1(x),S^{1}_2(+)}=h_{B_1(x),+-++},\\
&&h_3=h_{B_1(x),\sigma_5^{(1)}}=h_{B_1(x),\sigma_6^{(1)}}=h_{B_1(x),\sigma_7^{(1)}}=h_{B_1(x),S^{2}_1(+)}=h_{B_1(x),+--+},\\
&&h_4=h_{B_1(x),\sigma_8^{(1)}}=h_{B_1(x),S^{3}_0(+)}=h_{B_1(x),+---},\\
&&h_5=h_{B_1(x),\sigma_9^{(1)}}=h_{B_1(x),S^{0}_3(-)}=h_{B_1(x),-+++},\\
&&h_6=h_{B_1(x),\sigma_{10}^{(1)}}=h_{B_1(x),\sigma_{11}^{(1)}}=h_{B_1(x),\sigma_{12}^{(1)}}=h_{B_1(x),S^{1}_2(-)}=h_{B_1(x),--++},\\
&&h_7=h_{B_1(x),\sigma_{13}^{(1)}}=h_{B_1(x),\sigma_{14}^{(1)}}=h_{B_1(x),\sigma_{15}^{(1)}}=h_{B_1(x),S^{2}_1(-)}=h_{B_1(x),---+},\\
&&h_8=h_{B_1(x),\sigma_{16}^{(1)}}=h_{B_1(x),S^{0}_3(-)}=h_{B_1(x),----}.
\end{eqnarray*}

\begin{figure}[!t]
\centering
\includegraphics[width=65mm]{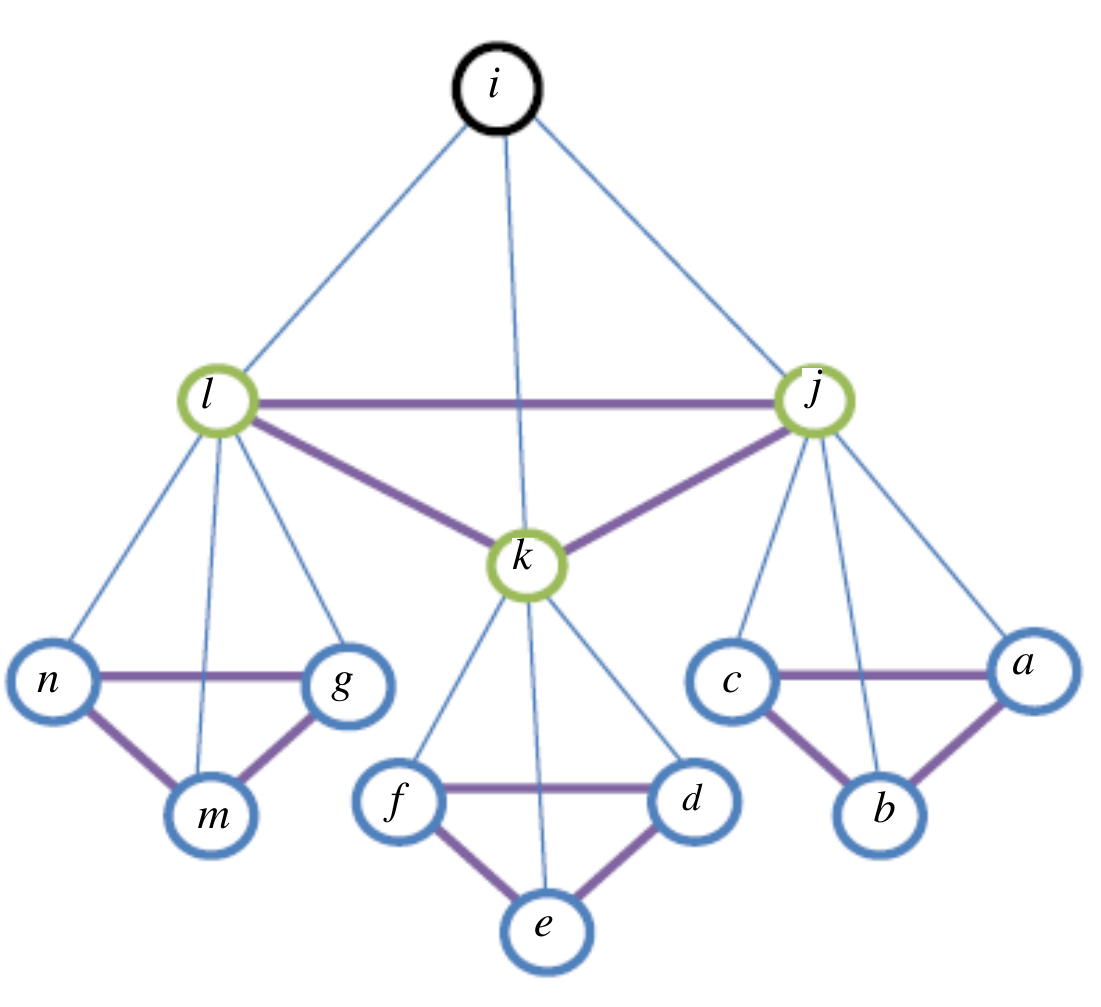}
\caption{(Colour online) Possible configurations of TCL with level
2 ($\Phi^{V_2^{(P)}}$). Schematic diagram to illustrate the
summation used in equation~\eqref{h-funct0}.}\label{configuration1}
\end{figure}

Therefore, we can define the vector-valued function $\textbf{h} :
V \rightarrow \mathbf{R}^8$ as follows:
\begin{eqnarray}\label{h-funct1a}
\textbf{h}(x)=(h_1,h_2, h_3, h_4, h_5, h_6, h_7, h_8).
\end{eqnarray}
By considering possible configurations  as in figure~\ref{configuration1} and from \eqref{compatible2a}, we can obtain
the following equation:
\begin{align}\label{h-funct0}
\exp\big[{ijkl\
h'_{B_{1}(x^{(0)}),ijkl}}\big]&=L_2\sum\limits_{\substack{a,b,c,d,f,e,\\g,m,n\in
\{-1,+1\}}}\Big(\exp\big\{J[j(a+b+c)+k(d+e+f)+l(g+m+n)]\big\}\nonumber\\ &\times
\exp\big\{J_\text{SL}[bc+a(b+c)+e f+d(e+f)+mn+g(m+n)]\big\}\nonumber\\
&\times \exp\big[J_\text{p} i(a+b+c+d+f+g+m+n+e)\big]\nonumber\\
&\times \exp\big[abcj \ h_{B_{1}(y^{(1)}),jabc}+defk \
h_{B_{1}(y^{(2)}),kdef}+glmn  \
h_{B_{1}(y^{(3)}),lgmn}\big]\Big).
\end{align}
Here, $i,j,k,l\in \{-1,+1\}$ and $y^{(1)},y^{(2)},y^{(3)}\in
S(x^{(0)})$ (see figure~\ref{configuration1}). Therefore,  we have
obtained an explicit formula for Gibbs measures with memory of
length 2 satisfying consistency conditions by means of equation
\eqref{h-funct0}.

From \eqref{h-funct0}, after long and complicated calculations we
get the following 8 equations:
\begin{align}\label{h-funct1}
\re^{h'_1}&=L_2\left(\frac{\re^{6J+6J_\text{p}+4J_\text{SL}+h_1}+3\re^{4J+4J_\text{p}-h_2}+3\re^{2J+2J_\text{p}+h_3}+\re^{4J_\text{SL}-h_4}}{\re^{3J+3J_\text{p}+J_\text{SL}}}\right)^3, \\
\re^{-h'_2}&=L_2\left(\frac{\re^{6J+6J_\text{p}+4J_\text{SL}+h_1}+3\re^{4J+4J_\text{p}-h_2}+3\re^{2J+2J_\text{p}+h_3}+\re^{4J_\text{SL}-h_4}}{\re^{3J+3J_\text{p}+J_\text{SL}}}\right)^2\nonumber %\\
\end{align}
\begin{align}
\label{h-funct2}
&\times\left(\frac{\re^{6J_\text{p}+4J_\text{SL}-h_5}+3\re^{2J+4J_\text{p}+h_6}+3\re^{4J+2J_\text{p}-h_7}+\re^{6J+4J_\text{SL}+h_8}}{\re^{3J+3J_\text{p}+J_\text{SL}}}\right),\\\label{h-funct3}
\re^{h'_3}&=L_2\left(\frac{\re^{6J+6J_\text{p}+4J_\text{SL}+h_1}+3\re^{4J+4J_\text{p}-h_2}+3\re^{2J+2J_\text{p}+h_3}+\re^{4J_\text{SL}-h_4}}{\re^{3J+3J_\text{p}+J_\text{SL}}}\right)\nonumber\\
&\times\left(\re^{6J_\text{p}+4J_\text{SL}-h_5}+3\re^{2J+4J_\text{p}+h_6}+3\re^{4J+2J_\text{p}-h_7}+\re^{6J+4J_\text{SL}+h_8}\right)^2,
\\\label{h-funct4}
\re^{-h'_4}&=L_2\left(\frac{\re^{6J_\text{p}+4J_\text{SL}-h_5}+3 \re^{2
J+4J_\text{p}+h_6}+3 \re^{4 J+2J_\text{p}-h_7}+\re^{6
J+4J_\text{SL}+h_8}}{\re^{3J+3J_\text{p}+J_\text{SL}}}\right)^3,\\
\label{h-funct5}
\re^{-h'_5}&=L_2\left(\frac{\re^{6J+4J_\text{SL}+h_1}+3\re^{4J+2J_\text{p}-h_2}+3\re^{2J+4J_\text{p}+h_3}+\re^{6J_\text{p}+4J_\text{SL}-h_4}}{\re^{3J+3J_\text{p}+J_\text{SL}}}\right)^3,\\\label{h-funct6}
\re^{h'_6}&=L_2\left(\frac{\re^{6J+4J_\text{SL}+h_1}+3\re^{4J+2J_\text{p}-h_2}+3\re^{2J+4J_\text{p}+h_3}+\re^{6J_\text{p}+4J_\text{SL}-h_4}}{\re^{3J+3J_\text{p}+J_\text{SL}}}\right)^2\nonumber\\
&\times
\left(\frac{\re^{4J_\text{SL}-h_5}+3\re^{2J+2J_\text{p}+h_6}+3\re^{4J+4J_\text{p}-h_7}+\re^{6J+6J_\text{p}+4J_\text{SL}+h_8}}{\re^{3J+3J_\text{p}+J_\text{SL}}}\right),\\\label{h-funct7}
\re^{-h'_7}&=L_2\left(\frac{\re^{6J+4J_\text{SL}+h_1}+3\re^{4J+2J_\text{p}-h_2}+3\re^{2J+4J_\text{p}+h_3}+\re^{6J_\text{p}+4J_\text{SL}-h_4}}{\re^{3J+3J_\text{p}+J_\text{SL}}}\right)\nonumber\\
&\times
\left(\frac{\re^{4J_\text{SL}-h_5}+3\re^{2J+2J_\text{p}+h_6}+3\re^{4J+4J_\text{p}-h_7}+\re^{6J+6J_\text{p}+4J_\text{SL}+h_8}}{\re^{3J+3J_\text{p}+J_\text{SL}}}\right)^2,\\\label{h-funct8}
\re^{h'_8}&=L_2 \left(\frac{\re^{4J_\text{SL}-h_5}+3 \re^{2 J+2J_\text{p}+h_6}+3 \re^{4
J+4J_\text{p}-h_7}+\re^{6 J+6
J_\text{p}+4J_\text{SL}+h_8}}{\re^{3J+3J_\text{p}+J_\text{SL}}}\right)^3.
\end{align}
\begin{rem}
From equations~\eqref{h-funct1}--\eqref{h-funct8}, one can easily
show that
\begin{eqnarray}\label{h-funct9} \re^{-3h'_2}=\re^{2h'_1-h'_4},\ \
\re^{3h'_3}=\re^{h'_1-2h'_4},\ \ \re^{3h'_6}=\re^{-2h'_5+h'_8},\ \
\re^{-3h'_7}=\re^{-h'_5+2h'_8}.
\end{eqnarray}
\end{rem}
\begin{rem} If the vector-valued function $\textbf{\textup{h}}(x)$ given in \eqref{h-funct1a} has the following form:
\begin{equation*}
\textbf{\textup{h}}(x)=\left(p,\frac{q-2p}{3},\frac{p-2q}{3},q,r,\frac{s-2r}{3},\frac{r-2s}{3},s\right),
\end{equation*}
then, the \textbf{consistency} condition \eqref{Gibbs-KCC} is
satisfied, where $p,q,r,s\in \mathbf{R}$.
\end{rem}

\section{Translation-invariant Gibbs measures (TIGMs) on a TCL}\label{Section-TIGMs}

In this section, we describe  a set of translation-invariant Gibbs
measures (TIGMs) associated with the model \eqref{Hm-C3}. Assume
that $a=\re^{J}$, $b=\re^{J_\text{p}}$ and $c=\re^{J_\text{SL}}$. By using the
equations \eqref{h-funct1}--\eqref{h-funct8}, we can take new
variables $u'_{i}=\exp[h_{B_1(x),S_{3-j}^j(\sigma(x))}]$ for $x\in
W_{n-1}$ and $u_{i}=\exp[h_{B_1(y),S_{3-j}^j(\sigma(y))}]$ for $y\in
S(x)$. Therefore, selecting the variables $u'_1$, $u'_4$ $u'_5$
and $u'_8$, we will obtain only the following equations:
\begin{align}\label{recurrence1aa}
u'_{1}&=L_2(ab)^{-9}c^{-3}\left(a^6b^6c^4u_1+3a^4b^4\frac{1}{u_2}+3a^2b^2u_3+\frac{c^4}{u_4}\right)^3,\\\label{recurrence2aa}
(u'_{4})^{-1}&=L_2(ab)^{-9}c^{-3}\left(\frac{b^6c^4}{u_5}+3a^2b^4u_6+\frac{3a^4b^2}{u_7}+a^6c^4u_8\right)^3,\\\label{recurrence3aa}
(u'_{5})^{-1}&=L_2(a
b)^{-9}c^{-3}\left(a^6c^4u_1+\frac{3a^4b^2}{u_2}+3a^2b^4u_3+\frac{a^6c^4}{u_4}\right)^3,\\\label{recurrence4aa}
u'_{8}&=L_2(ab)^{-9}c^{-3}\left[\frac{c^4}{u_5}+3(ab)^2u_6+\frac{3(ab)^4}{u_7}+(ab)^6c^4u_8\right]^3.
\end{align}
Substitution ${{u}_{i}}\to v_{i}^{3}$. From \eqref{h-funct9}, it
is clear that $v_2=({v_1^2}/{v_4})^{1/3}$ and
$v_3=({v_1}/{v_4^2})^{1/3}$. Similarly,
$v_6=({v_8}/{v_5^2})^{1/3}$ and
$v_7=({v_8^2}/{v_5})^{1/3}$.

Thus, we can derive the following recursion relation
\begin{align}\label{invar-2a}
v_{1}'&=\sqrt[3]{L_2}\frac{(a b)^6c^4(v_1 v_4){}^3+3(a
b)^4(v_1v_4)^2+3(a b)^2(v_1v_4)+c^4}{(a
b)^{3}cv_4^{3}}\,,\\\label{invar-2b}
(v_4')^{-1}&=\sqrt[3]{L_2}\frac{b^6c^4+3a^2b^4(v_5v_8)+3a^4b^2(v_5v_8){}^2+c^4a^6(v_5v_8)^3}{(ab)^{3}cv_5^{3}}\,,\\\label{invar-2c}
(v_5')^{-1}&=\sqrt[3]{L_2}\frac{a^6c^4(v_1v_4){}^3+3a^4b^2(v_1v_4){}^2+3a^2b^4(v_1v_4)+b^6c^4}{(a
b)^{3}c v_4^{3}}\,,\\\label{invar-2d}
 v_{8}'&=\sqrt[3]{L_2}\frac{c^4+3(ab)^2(v_5v_8)+3(ab)^4(v_5v_8){}^2+(ab)^6c^4(v_5v_8){}^3}{(ab)^3cv_5^3}.
\end{align}
Now we are going to investigate the derived system
\eqref{invar-2a}--\eqref{invar-2d}. To do this, let us consider the
following operator
\begin{eqnarray}\label{operator}
F:(v_{1},v_{4},v_{5},v_{8})\in \mathbf{R}_{+}^{4}\rightarrow
(v_{1}',v_{4}',v_{5}',v_{8}')\in \mathbf{R}_{+}^{4}.
\end{eqnarray}
It is generally a difficult problem to determine all invariant
sets according to operator $F$. Now to explain this situation, we
can write one of the invariant sets of the operator $F$ as
\[
\Upsilon:=\{(v_{1},v_{4},v_{5},v_{8})\in
\mathbf{R}_{+}^{4}:v_1=v_5 ,v_4=v_8\}.
\]
Note that one can show that the set $\Upsilon$ is invariant with
respect to the operator $F$, i.e., $F(\Upsilon)\subseteq
\Upsilon$. We can determine the invariant subsets of this
operator, which are used to describe the Gibbs distributions. The
equations corresponding to the restrictions of the operator $F$ to
all invariant sets are very cumbersome. Their solutions can be
determined using a computer, but this lies outside our circle of
interests. We note that the relation between the solutions of the
equations and the Gibbs distributions is determined by relations
\eqref{Gibbs-KCC} and \eqref{compatible2a}. The restriction of the
operator $F$ to the set $\Upsilon$ gives  some known of
Gibbs distributions. The restriction to $\Upsilon$ leads to new
Gibbs distributions.

Divide equation \eqref{invar-2a} by \eqref{invar-2c} and similarly
divide equation \eqref{invar-2d} by \eqref{invar-2b}. We have
\begin{align}\label{invar-21a}
v_{1}'v_5'&=\frac{(a b)^6c^4(v_1
v_4)^3+3(ab)^4(v_1v_4)^2+3(a
b)^2(v_1v_4)+c^4}{a^6c^4(v_1v_4)^3+3a^4b^2(v_1v_4)^2+3a^2b^4(v_1v_4)+b^6c^4}\,,\\\label{invar-21b}
v_4'v_{8}'&=\frac{c^4+3(ab)^2(v_5v_8)+3(ab)^4(v_5v_8)^2+(ab)^6c^4(v_5v_8)^3}
{b^6c^4+3a^2b^4(v_5v_8)+3a^4b^2(v_5v_8){}^2+c^4a^6(v_5v_8)^3}.
\end{align}

\section{Phase translations}\label{Phase translations}
Note that if there is more than one positive fixed point of the
operator \eqref{operator}, then there is more than one Gibbs
measure corresponding to these positive fixed points. One says
that a phase transition occurs for the Ising model, if the system
of equations \eqref{invar-2a}--\eqref{invar-2d} has more than one
solution \cite{G,Preston,GHRR}. The number of the solutions of
equations \eqref{invar-21a} and \eqref{invar-21b} depends on the
coupling constants and the parameter
$\beta =1/T$. 
If it is possible to find an exact value of temperature $T_\text{cr}$
such that a phase transition occurs for all $T< T_\text{cr}$, then
$T_\text{cr}$ is called a critical value of temperature.

In this paper, we will only examine the following situation.

From \eqref{invar-21a} and \eqref{invar-21b}, if we suppose
$v_1=v_5=v_4=v_8=\sqrt{x}$, then we have
\begin{equation}\label{TIGM-1a}
x'=f(x):=\frac{c^4+3 a^2 b^2 x+3 a^4 b^4 x^2+a^6 b^6
c^4x^3}{c^4b^6+3 a^2 b^4 x+3 a^4 b^2 x^2+a^6 c^4x^3}.
\end{equation}
Therefore, the set $\wp: =\{(v_{1},v_{4},v_{5},v_{8})\in
\mathbf{R}_{+}^{4}:v_1=v_5 =v_4=v_8=\sqrt{x}\}$ is invariant with
respect to the operator $F$. The restriction of the operator $F$
to the set $\wp$ is denoted by the respective symbol $F|_{\wp}=f$.
Where we assume that $F|_{\wp}=f$.

In order to investigate the phase transition of the model, we will
analyze the positive fixed points of the rational function $f$
with real coefficients as a dynamical system defined in
\eqref{TIGM-1a}.

The zeros of equation $x=f(x)$ are the zeros of equation
\begin{eqnarray}\label{roots1a}
p_4(x)=a^6 x^4+Bx^3-Cx^2+D x-1=0,
\end{eqnarray}
where $B=c^{-4}a^4 b^2 (a^2 b^4 c^4-3)$, 
$C=3c^{-4}a^2b^4(1-a^2)$, $D=c^{-4}b^2 (3 a^2-b^4
c^4)$ and $c^{4}<3$.

By using Descartes' rule of signs to find the zeroes of a
polynomial, we can determine the number of real solutions to
equation \eqref{roots1a}. We can count the number of sign changes.
For example, if $B>0$, $C>0$ and $D>0$, then there are three sign
changes in the ``positive'' case. This number ``three'' is the maximum
number of possible positive zeroes ($x$-intercepts) for the
polynomial $p_4(x)$. In this case, there are either 3 or 1
positive roots.

Now let us look at $p_4(-x)$ for the case if $B>0$, $C>0$ and $D>0$
(that is, having changed the sign on $x$, so this is the
``negative'' case). There is only one sign change in this ``negative''
case, so there is exactly one negative root. For the other cases,
similarly, we can estimate the possible number of positive and
negative roots of the polynomial $p_4(x)$ (table~\ref{Table}).

\begin{table}[!b]
  \centering
  \caption{The table of possible positive or negative roots of the polynomial $p_4(x)=c^4-B x+C x^2-D x^3-a^6 c^4 x^4$. }\label{Table}
  \vspace{2ex}
  \begin{tabular}{|c|c|c|c|c|}
    \hline\hline
    B & C & D & Positive roots & Negative roots\\\hline\hline
    + & + & + & 3,1 & 1 \\\hline
    + & + & - & 3,1 & 1\\\hline
    + & - & + & 1 & 3,1 \\\hline
    + & - & - & 3,1 & 1 \\\hline
    - & + & + & 1 & 3,1 \\\hline
    - & + & - & 1 & 3,1 \\\hline
    - & - & + & 1 & 3,1 \\\hline
    - & - & - & 3,1 & 1 \\
    \hline\hline
  \end{tabular}
\end{table}

Thus, we have two critical temperatures $T_\text{cr}^{*}=\frac{-J+2
(J_\text{p}+J_\text{LS})}{\ln\sqrt{3}}$ and $T_\text{cr}^{**}=\frac{J+2
(J_\text{p}+J_\text{LS})}{\ln\sqrt{3}}$.

For the antiferromagnetic Ising model ($J<0$) if we assume
\[\frac{-J+2 (J_\text{p}+J_\text{LS})}{\ln\sqrt{3}}<T<\frac{J+2
(J_\text{p}+J_\text{LS})}{\ln\sqrt{3}}
\]
then, there exist three positive fixed points of $f$.

Then, taking the first and the second derivatives of the function
$g$, we have
\[
f'(x)=\frac{3 a^2 \left(b^4-1\right) A(x,a,b,c)}{\left(b^2+a^2
x\right)^2 \left(b^4 c^4+3 a^2 b^2 x-a^2 b^2 c^4 x+a^4 c^4
x^2\right)^2}\,,
\]
where \begin{align*} A(x,a,b,c)&=b^4 c^4+2 a^2 b^2 c^4(1+b^4)x
+a^4 (3 b^4+c^8+b^4 c^8+b^8 c^8) x^2\\
&+2 a^6 b^2
c^4(1+b^4) x^3+a^8 b^4 c^4 x^4).
\end{align*}
If $b^4<1$ (with $x \geqslant0$), then $f$ is decreasing and there can
only be one solution of $f(x)=x,$ where it is obvious that
$A(x,a,b,c)>0$. Thus, we can restrict ourselves to the case in
which $b^4>1.$ That is, we will assume that $\frac{J_\text{p}}{T}>0$. 

\begin{thm}\label{theorem-main} For the Ising model on the TCL of
order 3, the following statements are true:
\begin{enumerate}
    \item If $T< T_\text{\textup{cr}}^{*}\,$, $T> T_\text{\textup{cr}}^{**}$ and  $\frac{J_\text{\textup{p}}}{T}<0$, then
there is a unique translation-invariant Gibbs measure $\mu_0$.
    \item  If $\frac{-J+2 (J_\text{\textup{p}}+J_\text{\textup{LS}})}{\ln\sqrt{3}}<T<\frac{J+2
(J_\text{\textup{p}}+J_\text{\textup{LS}})}{\ln\sqrt{3}}$, then there are 3
translation-invariant Gibbs measures $\mu_{-},\mu_{0},\mu_{+}$
indicating a phase transition ($\mu_{0}$ is called disordered
Gibbs measure). Moreover, $\mu_{-},\mu_{+}$ are extreme.
\end{enumerate}
\end{thm}

\subsection{An example indicating a phase transition}\label{An illustrative example}

By using an elementary analysis, we can obtain the fixed points of
the function $f$ given in \eqref{TIGM-1a} by finding real roots of
equation \eqref{roots1a}. Thus, we need to identify all the roots
of the polynomial \eqref{roots1a} of degree 4. Previously, a
documented analysis has solved these equations, which we will not
show here due to the complicated nature of formulas and
coefficients \cite{Wolfram}. Nonetheless, we have manipulated the
polynomial equation via Mathematica \cite{Wolfram}. Here, we will
only deal with positive fixed points, because of the positivity of
exponential functions.

Recall that the set of the fixed points of the function $f$ is
defined by
\[
\Fix(f)=\{x\in \mathbf{R}:f(x)=x\}.
\]
According to the size of the derivative, the fixed points are
classified as \begin{itemize}
    \item unstable\ if $|f'|>1$,
    \item neutral \ if $|f'|=1$,
    \item stable \ if $|f'|<1$,
    \item superstable\ if $|f'|=0.$
\end{itemize}

\begin{figure} [!t]
\centering
\includegraphics[width=65mm]{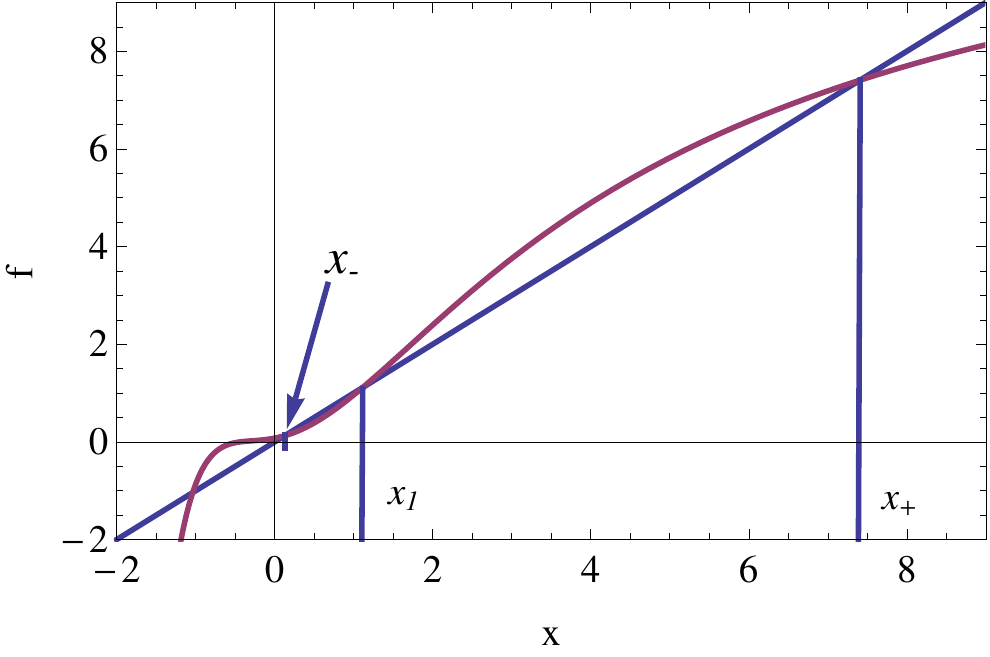}
\caption{(Colour online) There exist three positive fixed points of
the function $f$ for $J = -1$, $J_\text{p}= 29$,  $J_\text{SL}= 5.3$, $T =
68$.}\label{3fp3roots}
\end{figure}

We have obtained 3 positive real roots for some parameters $J$,
$J_\text{p}$ and $J_\text{SL}$ (coupling constants) and temperature $T$. For
example, in figure~\ref{3fp3roots}, we have manipulated that there
are 3 positive fixed points of the function \eqref{TIGM-1a} for $J
= -1$, $J_\text{p}= 29$, $J_\text{SL}= 5.3$, $T = 68$. As a result, there are
three translation-invariant Gibbs measures associated with the
positive fixed points. Therefore, for $J = -1$, $J_\text{p}= 29$, $J_\text{SL}=
5.3$, $T = 68$, the phase transitions occur.

For $J = -1 $, $J_\text{p}= 29$,  $J_\text{SL}= 5.3$, $T = 68$, there exist four fixed
points of the function $f$ obtained in \eqref{TIGM-1a} as follows
\[
\Fix(f)=\{ -1.0376,0.127421,1.11525, 7.40762\}.
\]

It is clear that $|f'(0.127421)|=0.470903<1$ and
$|f'(7.40762)|=0.520525<1$. Therefore, $x_-=0.127421$ and
$x_+=7.40762$ are stable fixed points of the function $f$.
The corresponding Gibbs measures are extreme ones. Also,
$|f'(1.11525)|=1.36756>1$, thus, $x_1=1.11525$ is unstable fixed
point of the function $f$ (see figure~\ref{3fp3roots}).

\begin{figure}[!t]
\centering
\includegraphics[width=65mm]{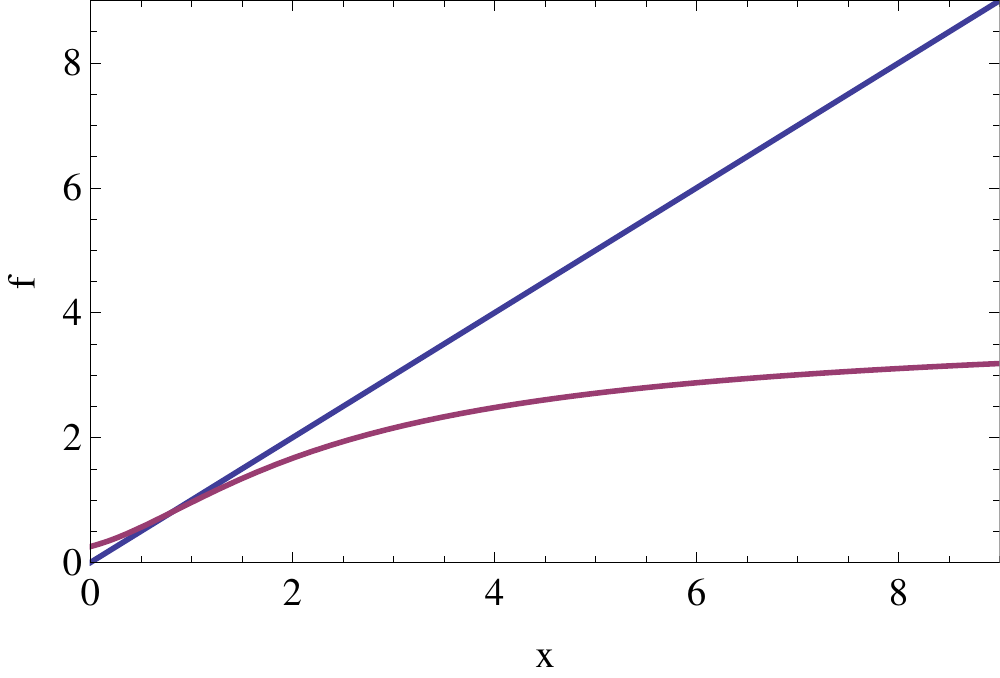}
\caption{(Colour online) There exists a unique positive fixed point
of the function $f$ for $J = -1$, $J_\text{p}= 10$,  $J_\text{SL}= 5.3$, $T =
44$.}\label{3fp1root}
\end{figure}

In figure~\ref{3fp1root}, there exists only a single positive fixed
point of the function \eqref{TIGM-1a} for $J = -1$, $J_\text{p}= 10$, $J_\text{SL}=
5.3$, $T = 44$. The set of the other fixed points are $\{-1.05633,
0.554978 -1.02241 \ri,0.554978 +1.02241 \ri,0.801718\}$. Therefore,
there is a unique Gibbs measure corresponding to the fixed point
$x=0.801718$ on the CL with parameters $J = -1$, $J_\text{p}= 10$, $J_\text{SL}=
5.3$, $T = 44$. Therefore, the phase transition does not occur for
$J = -1$, $J_\text{p}= 10$,  $J_\text{SL}= 5.3$, $T = 44$.

Note that an attractive fixed point of a function $f$ is a fixed
point $x_+$ of $f$ such that for any value of $x$ in a domain that
is close enough to $x_0$, the iterated function sequence $x, f(x),
f(f(x)), f(f(f(x))), \ldots$ converges to $x_+$. An attractive
fixed point is said to be a stable fixed point if it is also
Lyapunov stable (see \cite{AkinT2011CMP} for details).

\begin{rem}
From theorem \ref{theorem-main}, one can say that the stable roots
describe extreme Gibbs distributions. Therefore, from the figure~\ref{3fp3roots}, we can conclude that the Gibbs measures $\mu_{-}$
and $\mu_{+}$ corresponding to the stable fixed points $x_{-}$ and
$x_{+}$ are extreme Gibbs distributions
\cite{Iof,NHSS,NHSS1,Rozikov}.
\end{rem}
\begin{rem}
We conclude that there are at most 3 translation-invariant Gibbs
measures corresponding to the positive real roots of the equation
\eqref{TIGM-1a}. Also, one can show that translation-invariant
Gibbs measures corresponding stable solutions are extreme.
\end{rem}

\section{Conclusions}\label{Conclusions}

When our model is compared with the model given in
\cite{Akin2017}, we can see the role of the competing coupling
$J_\text{SL}$ which represents the same level nearest neighbour on the phase
transition phenomenon. If we take $J_\text{SL}=0$, then the equation
\label{TIGM-1} is the same as the dynamical system in \cite{Akin2017}.
The recurrence equations obtained in the present paper totally
differ from \cite{Akin2016,Akin2017,Akin2017a,MAKfree2017}. Note
that for the Ising model associated with the Hamiltonian
$\eqref{Hm-C3}$ on the chandelier lattices of order $k$, in
contrast to the symmetry of arbitrary order Cayley tree
\cite{UGAT,Akin2017a}, if $k>3$, then the chandelier lattice of
order $k$ is not symmetry. Therefore, in order to construct the
recurrence equations associated with the given Hamiltonian
\eqref{Hm-C3} for $k>3$ is much more difficult.

To describe the set of all the corresponding Gibbs distributions
is one of the main problems for the given Hamiltonian
\cite{Nazarov-Rozikov}. However, the attempts to completely describe
this set have not been accomplished until now, even for rather
simple Hamiltonians. An exact description of all positive fixed
points of the operator $F$ given in \eqref{operator} is rather
tricky. Therefore, under some assumptions, description of the
translation-invariant Gibbs measures for the model has been given.
We have also shown that for some parameter values of the model
there is a phase transition. We state some unsolved problems that
turned out to be rather complicated and require further
consideration:
\begin{enumerate}
    \item Do any other invariant sets of the operator $F$ exist?
    \item Do positive fixed points of the operator $F$ exist outside the invariant sets?
\end{enumerate}
Ganikhodjaev and  Rozikov \cite{GR} have given a complete
description of periodic Gibbs measures for the Ising model, i.e., a
characterization of such measures with respect to any normal
subgroup of finite index in $G_k$.  Ak\i n et al. \cite{AGUT} have
studied the periodic extreme Gibbs measures with memory length~2
of Vannimenus model. Description  of the periodic (non
transition-invariant) Gibbs measures with a memory of length 2 on
the chandelier lattice remains an open problem.

In \cite{GRRR,MAKfree2017}, the authors have presented, for the
Ising model on the Cayley tree, some explicit formulae of the free
energies (and entropies) according to boundary conditions (b.c.).
By applying the general formulae to various known boundary
conditions on arbitrary order chandelier-lattices, we plan to
obtain some explicit formula of free energy and relative entropy
corresponding to the boundary conditions in our future work.

We think that the present paper is of certain interest for the
statistical physics community due to an interesting application of
the \textbf{chandelier-lattices} model to real problems.

\ukrainianpart

\title{Ґібсові міри моделі Ізінга з конкурентними взаємодіями на трикутній  люстровій гратці}
\author{Г. Акин}
\address{Вулиця Сейгун Атуф Кансу, 1164. Сокак, 9/4, TR06105, Санкая, Анкара, Туреччина}

\makeukrtitle

\begin{abstract}
	В цій статті ми розглядаємо модель Ізінга з трьома конкурентними взаємодіями на трикутній люстровій гратці. Описано існування, єдиність і неєдиність трансляційно інваріантних Ґібсових мір, пов'язаних з моделлю Ізінга. Отримано явну формулу для Ґібсових мір з пам'яттю довжиною 2, що задовільняють умови консистентності. 
	Строго доведено, що дана  модель проявляє фазові переходи лише для даних констант зв'язку. В результаті застосування даного підходу вирішено проблему дихотомії між альтернативними розв'язками Гамільтонових моделей на трикутних  люстрових гратках. Нарешті, показано два числових приклади, що ілюструють корисність і ефективність запропонованих теоретичних результатів.

\keywords  люстрові гратки, Ґібсові міри, модель Ізінга, фазовий перехід
\end{abstract}

\end{document}